\documentclass[journal]{IEEEtran}
\PassOptionsToPackage{ruled}{algorithm2e}
\usepackage[T1]{fontenc}
\usepackage[utf8]{inputenc}
\usepackage{color}
\usepackage{verbatim}
\usepackage{refstyle}
\usepackage{algorithm2e}
\usepackage{amsmath}
\usepackage{amsthm}
\usepackage{amssymb}
\usepackage{graphicx}
\usepackage[unicode=true,
 bookmarks=true,bookmarksnumbered=true,bookmarksopen=true,bookmarksopenlevel=1,
 breaklinks=false,pdfborder={0 0 0},pdfborderstyle={},backref=false,colorlinks=false]
 {hyperref}
\hypersetup{pdftitle={Your Title},
 pdfauthor={Your Name},
 pdfpagelayout=OneColumn, pdfnewwindow=true, pdfstartview=XYZ, plainpages=false}

\makeatletter

\AtBeginDocument{\providecommand\algref[1]{\ref{alg:#1}}}
\RS@ifundefined{subsecref}
  {\newref{subsec}{name = \RSsectxt}}
  {}
\RS@ifundefined{thmref}
  {\def\RSthmtxt{theorem~}\newref{thm}{name = \RSthmtxt}}
  {}
\RS@ifundefined{lemref}
  {\def\RSlemtxt{lemma~}\newref{lem}{name = \RSlemtxt}}
  {}

\theoremstyle{plain}
\newtheorem{thm}{\protect\theoremname}
\theoremstyle{remark}
\newtheorem{rem}[thm]{\protect\remarkname}

\usepackage[caption=false,font=footnotesize]{subfig}
\usepackage{cite}

\DeclareMathOperator{\maximize}{max.}

\DeclareMathOperator{\st}{s.t.}

\DeclareMathOperator{\tr}{Tr}

\newref{alg}{refcmd = \textbf{Algorithm~\ref{#1}}}

\usepackage{cite}
\usepackage{acronym}
\AtBeginDocument{\acrodef{6G}{sixth-generation}
\acrodef{AO}{alternating optimization}
\acrodef{AoA}{angle of arrival}
\acrodef{AoD}{angle of departure}
\acrodef{APGM}{alternating projected gradient method}
\acrodef{APM}{accelerated proximal gradient method}
\acrodef{AP}{access point}
\acrodef{ASP}{antenna separation product}
\acrodef{AWGN}{additive white Gaussian noise}
\acrodef{BC}{broadcast channel}
\acrodef{BCM}{block coordinate maximization}
\acrodef{BEP}{bit error probability}
\acrodef{BER}{bit error rate}
\acrodef{BF-MIMO}[BF\mbox{-}MIMO]{beamforming MIMO}
\acrodef{BF}{beamforming}
\acrodef{BS}{base station}
\acrodef{bpcu}{bits per channel use}
\acrodef{CP}{cyclic prefix}
\acrodef{CPU}{central processing unit}
\acrodef{CR}{communication rate}
\acrodef{CSI}{channel state information}
\acrodef{CSIR}{channel state information at RX}
\acrodef{SSK}{space shift keying}
\acrodef{CRLB}{Cramer-Rao lower bound}
\acrodef{CSIT}{channel state information at TX}
\acrodef{DCMC}{discrete\mbox{-}input continuous\mbox{-}output memoryless channel}
\acrodef{DFT}{discrete Fourier transform}
\acrodef{DL-TR-GSM}{dual-layered transmit-receive \acl{GSM}}
\acrodef{DLT}{dual-layered transmission}
\acrodef{DMA}{dynamic metasurface antenna}
\acrodef{DOA}{direction of arrival}
\acrodef{DoF}{degrees of freedom}
\acrodef{DNN}{deep neural network}
\acrodef{DPC}{dirty paper coding}
\acrodef{DRL}{deep reinforcement learning}
\acrodef{EE}{energy efficiency}
\acrodef{EGC}{equal gain combining}
\acrodef{EM}{electromagnetic}
\acrodef{EVD}{eigenvalue decomposition}
\acrodef{FPGA}{field programmable gate array}
\acrodef{FSPL}{free space path loss}
\acrodef{FFT}{fast Fourier transform}
\acrodef{FDE}{frequency domain equalization}
\acrodef{GRSM}{generalized \acl{RSM}}
\acrodef{GSM}{generalized \acl{SM}}
\acrodef{HMIMO}{holographic MIMO}
\acrodef{IA}{inner approximation}
\acrodef{IFFT}{invserse fast Fourier transform}
\acrodef{ICI}{inter-channel interference}
\acrodef{iid}[i.i.d.]{independent and identically distributed}
\acrodef{IMT}{International Mobile Telecommunications}
\acrodef{IQ}{in\mbox{-}phase and quadrature}
\acrodef{ISAC}{integrated sensing and communication}
\acrodef{ISI}{intersymbol interference}
\acrodef{ISI-free}[ISI\mbox{-}free]{intersymbol interference free}
\acrodef{LIS}{large intelligent surface}
\acrodef{LOS}{line\mbox{-}of\mbox{-}sight}
\acrodef{KKT}{Karush\mbox{-}Kuhn\mbox{-}Tucker} 
\acrodef{MAC}{multiple-access channel}
\acrodef{mmWave}{millimeter-wave}
\acrodef{MI}{mutual information}
\acrodef{MIMO}{multiple\mbox{-}input multiple\mbox{-}output}
\acrodef{mMIMO}{massive MIMO}
\acrodef{MISO}{multiple\mbox{-}input single\mbox{-}output}
\acrodef{ML}{maximum likelihood}
\acrodef{MRC}{maximal ratio combining}
\acrodef{MMSE}{minimum mean square error}
\acrodef{MU-TR-GSM}{multiuser transmit-receive  \acl{GSM} }
\acrodef{NCSIT}{no channel state information at TX}
\acrodef{NLOS}{non\mbox{-}\acs{LOS}} 
\acrodef{NOMA}{non-orthogonal multiple access}
\acrodef{OFDM}{orthogonal frequency division multiplexing}
\acrodef{OFDMA}{orthogonal frequency division multiple access}
\acrodef{OMP}{orthogonal matching pursuit}
\acrodef{OTFS}{orthogonal time frequency space}
\acrodef{umMIMO}{ultra-massive MIMO}
\acrodef{PA}{power amplifier}
\acrodef{PAE}{power added efficiency}
\acrodef{PAPR}{peak\mbox{-}to\mbox{-}average power ratio}
\acrodef{PDF}{probability density function}
\acrodef{PEP}{pairwise error probability}
\acrodefplural{PEP}{pairwise error probabilities}
\acrodef{PGM}{projected gradient method}
\acrodef{PMP}{probability mass function}
\acrodef{PSM}{precoding-aided spatial modulation}
\acrodef{QSM}{quadrature spatial modulation}
\acrodef{RC}{reorganization computation}
\acrodef{RCS}{radar cross section}
\acrodef{RF}{radio frequency}
\acrodef{RHS}{right-hand side}
\acrodef{RIS}{reconfigurable intelligent surface}
\acrodef{RSM}{receive spatial modulation}
\acrodef{RX}{receiver}
\acrodef{SE}{spectral efficiency}
\acrodef{SEP}{symbol error probability}
\acrodef{SER}{symbol error rate}
\acrodef{SIC}{successive interference cancellation}
\acrodef{SIM}{stacked intelligent metasurface}
\acrodef{SINR}{signal-to-interference-plus-noise ratio}
\acrodef{SISO}{single-input single-output}
\acrodef{SM}{spatial modulation}
\acrodef{SMX-MIMO}[SMX\mbox{-}MIMO]{spatial multiplexing MIMO}
\acrodef{SMX}{spatial multiplexing}
\acrodef{SNR}{signal-to-noise ratio}
\acrodef{SC}{single carrier}
\acrodef{SCA}{successive convex approximation}
\acrodef{SVD}{singular value decomposition}
\acrodef{SPST}{single pole single-throw}
\acrodef{SR}{sensing rate}
\acrodef{SU}{secondary user}
\acrodef{TDE}{time domain equalization}
\acrodef{THz}{teraherz}
\acrodef{TX}{transmitter}
\acrodef{ULA}{uniform linear array}
\acrodef{URA}{uniform rectangular array}
\acrodef{VGA}{variable gain amplifier}
\acrodef{wrt}[w.r.t.]{with respect to}
\acrodef{ZF}{zero-forcing}
\acrodef{ZMCG}{zero-mean complex Gaussian}

}
\ifdefined\showcaptionsetup
 \PassOptionsToPackage{caption=false}{subfig}
\fi
\usepackage{subfig}
\makeatother

\providecommand{\remarkname}{Remark}
\providecommand{\theoremname}{Theorem}

\begin{document}
\title{Sensing Rate Optimization for Multi-Band Cooperative ISAC Systems}
\author{Nemanja Stefan Perovi\'c, Mark F. Flanagan,~\IEEEmembership{Senior Member,~IEEE,}
and~Le-Nam Tran,~\IEEEmembership{Senior Member,~IEEE}\thanks{Nemanja Stefan~Perovi\'c was with Universit\'e Paris-Saclay, CNRS,
CentraleSup\'elec, Laboratoire des Signaux et Syst\`emes, 3 Rue
Joliot-Curie, 91192 Gif-sur-Yvette, France. He is now with the Institute
of Communications Engineering, National Sun Yat-sen University, Kaohsiung
80424, Taiwan, R.O.C. (Email: n.s.perovic@mail.nsysu.edu.tw).}\thanks{Mark F. Flanagan and Le-Nam~Tran are with the School of Electrical
and Electronic Engineering, University College Dublin, Belfield, Dublin
4, D04~V1W8, Ireland (Email: mark.flanagan@ieee.org, nam.tran@ucd.ie).}\vspace{-0.5cm}}
\maketitle
\begin{abstract}
Integrated sensing and communication (ISAC) has been recognized as
one of the key technologies for future wireless networks, which potentially
need to operate in multiple frequency bands to satisfy ever-increasing
demands for both communication and sensing services. Motivated by
this, we consider the sum \ac{SR} optimization for a cooperative
ISAC system with linear precoding, where each \ac{BS} works in a
different frequency band. With this aim, we propose an optimization
algorithm based on the semi-definite rank relaxation that introduces
covariance matrices as optimization variables, and we apply the \ac{IA}
method to deal with the nonconvexity of the resulting problem. Simulation
results show that the proposed algorithm increases the SR by approximately
25\,\% and 40\,\% compared to the case of equal power distribution
in a cooperative ISAC system with two and three BSs, respectively.
Additionally, the algorithm converges in only a few iterations, while
its most beneficial implementation scenario is in the low power regime.\acresetall{}
\end{abstract}

\begin{IEEEkeywords}
Cooperative, \ac{CR}, \ac{ISAC}, multi-band, \ac{SR}. \acresetall{}
\end{IEEEkeywords}

\section{Introduction}

\bstctlcite{BSTcontrol}In addition to conventional communication
services, future sixth-generation (6G) communication systems will
need to provide high-precision sensing as a new necessary functionality
underpinning different applications (e.g., augmented reality, digital
twins). A promising technology for the implementation of this functionality
with minimal additional resource usage is that of \ac{ISAC} \cite{liu2022integrated}.
This refers to a design paradigm in which sensing and communication
systems are integrated to efficiently utilize the shared spectrum
and hardware resources, while offering mutual benefits\textcolor{black}{{}
\cite{magbool2024multi}}. As such, \ac{ISAC} is expected to be one
of the key technologies in \ac{6G} networks, which offers flexible
trade-offs between the two functionalities across various use cases.

With the rise of high-data rate applications, wireless traffic demand
is sharply increasing. This leads to spectrum congestion in the \ac{RF}
band, which is insufficient for higher rate services. To meet these
demands, all available spectrum, including sub-6 GHz, \ac{mmWave}
and \ac{THz} bands, must be fully utilized \cite{wang2023road}.
Furthermore, high-frequency technology also enables advanced sensing
and localization with centimeter-level accuracy. Hence, frequency
resource allocation is crucial for supporting both communication and
sensing demands. Ultimately, future wireless networks will operate
simultaneously in different frequency bands and will also need to
possess cooperative capabilities. This fact motivates the development
of cooperative ISAC systems that simultaneously operate in more than
one frequency band.

Recent studies have explored ISAC in cooperative communication systems.
A joint \ac{BS} mode selection, transmit beamforming, and receive
filter design for cooperative cell-free ISAC networks, where multiple
\acp{BS} cooperatively serve communication users and detect targets,
were studied in \cite{liu2024cooperativecellfree}. In \cite{meng2024cooperative},
stochastic geometry was used to analyze scaling laws in cooperative
ISAC networks,\textcolor{black}{{} and it was shown that the sensing
performance increases significantly with the number of ISAC transceivers,
especially when all ISAC transceivers are equidistant from the target.}
In \cite{pucci2024cooperative}, a \ac{ML} framework for target localization
was developed, deriving a cooperative position error bound (PEB) for
the networked \ac{OTFS}-based system which serves as a lower bound
on the positioning accuracy. Performance analysis and optimal power
allocation in a cooperative ISAC system with a micro and a macro BS,
where the micro BS performs sensing and acts as a full-duplex relay,
were presented in \cite{liu2022performance}.

In contrast to existing studies on cooperative ISAC in a single frequency
band, multi-band cooperative ISAC remains largely unexplored. In \cite{liu2024target},
a relatively complex method for fusion of \ac{OFDM} sensing signals
from different bands was proposed which achieved high-accuracy target
localization, requiring dual-band processing at both BSs. A similar
approach but for single-BS ISAC in \cite{liu2024carrier} improved
the sensing \ac{CRLB} and communication \ac{MI}, though it did not
account for propagation loss differences between frequency bands.

Against this background, the contributions of this paper are listed
as follows:
\begin{itemize}
\item We extend the \ac{MI} framework developed in \cite{li2024framework}
to a multi-band cooperative ISAC system, where each BS operates in
a different frequency band. To maximize the sum \ac{SR} for this
system, we formulate a joint optimization problem of the transmit
precoding matrices. For this purpose, we adopt linear precoding. To
maintain user fairness, each user has a guaranteed minimum \ac{CR}.
\item We solve the considered problem by employing the semi-definite rank
relaxation method, which aims to find the covariance matrices of the
corresponding precoding matrices. The \ac{IA} method is applied to
deal with the nonconvexity of the resulting problem.
\item We demonstrate through simulations that the proposed algorithm \emph{increases
the SR by approximately 25\,\% and 40\,\%} compared to the case
of equal power distribution in a cooperative ISAC system with two
and three BSs, respectively. Additionally, the algorithm converges
in a couple of iterations. Moreover, we show that the operating frequency
and the distance from the sensing target have a significant impact
on the SR. Lastly, we conclude that the most preferred application
scenario for our proposed system model is in the low power regime.
\end{itemize}
\textcolor{black}{\emph{Notation}}\textcolor{black}{: Bold lower and
upper case letters represent vectors and matrices, respectively. $\mathbf{I}_{x}$
is the unit matrix of size $x\times x$. $\text{Tr}(\mathbf{X})$,
$\text{vec}(\mathbf{X})$, $\det(\mathbf{X})$ and $\text{rank}(\mathbf{X})$
denote the trace, vectorization, determinant and rank of matrix $\mathbf{X}$,
respectively. $\ln(\cdot)$ is the natural logarithm and $\otimes$
denotes the Kronecker product. $(\cdot)^{*}$ denotes complex-conjugate.
$(\cdot)^{T}$ and $(\cdot)^{H}$ represent transpose and Hermitian
transpose, respectively.}

\section{System Model and Problem Formulation}

We consider a system with $B$ BSs, each being equipped with $N_{t}$
transmit and $N_{r}$ receive antennas, simultaneously serving $K$
communication users and performing target sensing. Each BS operates
on a different frequency band. A BS $b\in\mathcal{B}\triangleq\{1,2,\ldots,B\}$
operates at center frequency $f^{(b)}$, corresponding wavelength
$\lambda^{(b)}$, and bandwidth $B^{(b)}$. Each user $k\in\mathcal{K}\triangleq\{1,2,\ldots,K\}$
has $N_{k}$ antennas. Data symbols of unit average power are transmitted
over the period of $L$ time slots and are denoted as $\mathbf{S}^{(b)}=\bigl[(\mathbf{S}_{1}^{(b)})^{T},(\mathbf{S}_{2}^{(b)})^{T},\ldots,(\mathbf{S}_{K}^{(b)})^{T}\bigr]^{T}$,
where $\mathbf{S}_{k}^{(b)}\in\mathbb{C}^{N_{k}\times L}$ represents
data symbols transmitted to user $k$ and $N_{\mathrm{tot}}=\sum_{k=1}^{K}N_{k}$.
For asymptotically large $L$, we have
\begin{equation}
\mathbf{S}^{(b)}(\mathbf{S}^{(b)})^{H}\approx L\mathbf{I}_{N_{\mathrm{tot}}}.\label{eq:Sapprox}
\end{equation}

To provide high-quality communication and sensing links, data for
the $k$-th user is precoded using $\mathbf{W}_{k}^{(b)}\in\mathbb{C}^{N_{t}\times N_{k}}$.
At low-frequency bands, fully digital precoding is applied, while
at high-frequency bands, a hybrid precoding scheme is employed with
$\mathbf{W}_{k}^{(b)}=\mathbf{W}_{RF,k}^{(b)}\mathbf{W}_{BB,k}^{(b)}$,
where $\mathbf{W}_{BB,k}^{(b)}\in\mathbb{C}^{N_{k}\times N_{k}}$
is the fully digital baseband precoder and $\mathbf{W}_{RF,k}^{(b)}\in\mathbb{C}^{N_{t}\times N_{k}}$
as the analog precoder implemented using RF phase shifters. Hence,
all of the elements of $\mathbf{W}_{RF,k}^{(b)}$ must satisfy 
\begin{equation}
|\mathbf{W}_{RF,k}^{(b)}(i,j)|=1,\quad i=1,2,\ldots,N_{t},j=1,2,\ldots,N_{k}.
\end{equation}
After precoding, the transmitted signal can be written as
\begin{equation}
\mathbf{X}^{(b)}=\mathbf{W}^{(b)}\mathbf{S}^{(b)}=\sum\nolimits_{k=1}^{K}\mathbf{W}_{k}^{(b)}\mathbf{S}_{k}^{(b)}\in\mathbb{C}^{N_{t}\times L}
\end{equation}
where $\mathbf{W}^{(b)}=\bigl[\mathbf{W}_{1}^{(b)},\mathbf{W}_{2}^{(b)},\ldots,\mathbf{W}_{K}^{(b)}\bigr]\in\mathbb{C}^{N_{t}\times N_{\text{tot}}}$.

\subsection{Communication Model}

For the frequency band $b$, the received signals of all users can
be stacked to form a matrix $\mathbf{Y}_{C}^{(b)}$, given by 
\begin{equation}
\mathbf{Y}_{C}^{(b)}=\mathbf{H}^{(b)}\mathbf{X}^{(b)}+\mathbf{N}^{(b)}\in\mathbb{C}^{N_{\mathrm{tot}}\times L}
\end{equation}
where $\mathbf{H}^{(b)}=\bigl[\mathbf{H}_{1}^{(b)T},\mathbf{H}_{2}^{(b)T},\cdots,\mathbf{H}_{K}^{(b)T}\bigr]^{T}\in\mathbb{C}^{N_{\mathrm{tot}}\times N_{t}}$
with $\mathbf{H}_{k}^{(b)}\in\mathbb{C}^{N_{k}\times N_{t}}$ being
the channel matrix between the $b$-th BS and the $k$-th user. $\mathbf{N}^{(b)}\in\mathbb{C}^{N_{\mathrm{tot}}\times L}$
is the noise matrix consisting of \ac{iid} elements that are distributed
according to $\mathcal{CN}(0,(\sigma^{(b)})^{2})$. Assuming a sparsely-scattered
channel model, the channel matrix can be expressed as{\small
\begin{equation}
\mathbf{H}_{k}^{(b)}=\sqrt{\frac{N_{t}N_{k}}{P^{(b)}}}\sum\nolimits_{p=1}^{P^{(b)}}\beta_{p,k}^{(b)}\mathbf{a}_{\mathrm{R}}(\theta_{p,k}^{(b)})\mathbf{a}_{\mathrm{T}}(\phi_{p,k}^{(b)})^{H}
\end{equation}
}where $P^{(b)}$ is the number of signal paths and $\beta_{p,k}^{(b)}$
is the gain of path $p$ which is distributed according to $\mathcal{CN}(0,F_{k}^{(b)})$.
In the above, $F_{k}^{(b)}$ is the free space path loss, calculated
as $F_{k}^{(b)}=(\lambda^{(b)}/4\pi d_{k}^{(b)})^{2}$, where $d_{k}^{(b)}$
is the distance between the BS $b$ and user $k$. Also, $\theta_{p,k}^{(b)}$
and $\phi_{p,k}^{(b)}$ denote the \ac{AoA} and \ac{AoD} for the
$p$-th path between the BS and user $k$. The transmit and receive
antenna array responses are given by $\mathbf{a}_{\mathrm{T}}(\theta_{t}^{(b)})=\frac{1}{\sqrt{N_{t}}}[1,e^{j\frac{2\pi s_{t}^{(b)}}{\lambda^{(b)}}\sin\theta_{p,k}^{(b)}},\cdots,e^{j(N_{t}-1)\frac{2\pi s_{t}^{(b)}}{\lambda^{(b)}}\sin\theta_{p,k}^{(b)}}]^{T}$
and $\mathbf{a}_{\mathrm{R}}(\theta_{r}^{(b)})=\frac{1}{\sqrt{N_{k}}}[1,e^{j\frac{2\pi s_{k}}{\lambda^{(b)}}\sin\phi_{p,k}^{(b)}},\cdots,e^{j(N_{k}-1)\frac{2\pi s_{k}}{\lambda^{(b)}}\sin\phi_{p,k}^{(b)}}]^{T}$,
respectively, where $s_{t}^{(b)}$ and $s_{k}$ are the spacing between
the adjacent transmit and receive antennas, respectively.

For a single frequency band with bandwidth $B^{(b)}$, the achievable
rate for user $k$ can be expressed as
\begin{multline}
R_{C,k}^{(b)}=B^{(b)}\ln\det\bigl[\mathbf{I}_{N_{k}}+\mathbf{H}_{k}^{(b)}\mathbf{W}_{k}^{(b)}\mathbf{W}_{k}^{(b)H}\mathbf{H}_{k}^{(b)H}\\
\!\!\times\!\!\bigl(\sum\nolimits_{i=1,i\neq k}^{K}\mathbf{H}_{k}^{(b)}\mathbf{W}_{i}^{(b)}\mathbf{W}_{i}^{(b)H}\mathbf{H}_{k}^{(b)H}+(\sigma^{(b)})^{2}\mathbf{I}_{N_{k}}\bigr)^{-1}\bigr].\label{eq:Rck}
\end{multline}

\subsection{Sensing Model}

For target detection, each BS utilizes collocated transmit and receive
antennas to perform monostatic sensing. We assume that the transmit
and receive antennas at the same BS are separated and sufficiently
isolated from one another, so that any self-interference can be ignored.
Therefore, the AoD and AoA for each sensing signal are the same. If
no scattering interference is present in the sensing channel, the
echo signal is given by{\small
\begin{equation}
\mathbf{Y}_{S}^{(b)}=\mathbf{G}^{(b)}\mathbf{X}^{(b)}+\mathbf{Z}^{(b)}=\mathbf{G}^{(b)}\mathbf{W}^{(b)}\mathbf{S}^{(b)}+\mathbf{Z}^{(b)}
\end{equation}
}where $\mathbf{G}^{(b)}\in\mathbb{C}^{N_{r}\times N_{t}}$ is the
target response matrix and $\mathbf{Z}^{(b)}\in\mathbb{C}^{N_{r}\times L}$
is the noise matrix whose elements are distributed as $\mathcal{CN}(0,(\sigma^{(b)})^{2})$.
This equation can be further reformulated as \cite{li2024framework}{\small
\begin{equation}
\mathbf{y}_{S}^{(b)}=\widetilde{\mathbf{W}}^{(b)}\mathbf{\widetilde{S}}^{(b)}\mathbf{g}^{(b)}+\mathbf{z}^{(b)}
\end{equation}
}where $\mathbf{\widetilde{S}}^{(b)}=\mathbf{I}_{N_{r}}\otimes(\mathbf{S}^{(b)})^{H}$,
$\mathbf{\widetilde{W}}^{(b)}=\mathbf{I}_{N_{r}}\otimes(\mathbf{W}^{(b)})^{H}$,
$\mathbf{y}_{S}^{(b)}=\text{vec}((\mathbf{Y}_{S}^{(b)})^{H})$, $\mathbf{g}^{(b)}=\text{vec}((\mathbf{G}^{(b)})^{H})$
and $\mathbf{z}^{(b)}=\text{vec}((\mathbf{Z}^{(b)})^{H})$.

For a point target, the response matrix and its covariance matrix
are given by{\small
\begin{align}
\mathbf{G}^{(b)} & =\alpha^{(b)}\sqrt{N_{t}N_{r}}\mathbf{a}_{\mathrm{R}}(\varphi{}^{(b)})\mathbf{a}_{\mathrm{T}}(\varphi{}^{(b)})^{H}\\
\mathbf{R}^{(b)} & =\mathbb{E}\{\mathbf{g}^{(b)}\mathbf{g}^{(b)H}\}=\gamma N_{t}N_{r}(\mathbf{a}_{\mathrm{R}}(\varphi{}^{(b)})^{*}\otimes\mathbf{a}_{\mathrm{T}}(\varphi{}^{(b)}))\nonumber \\
 & \quad\quad\times(\mathbf{a}_{\mathrm{R}}(\varphi{}^{(b)})^{*}\otimes\mathbf{a}_{\mathrm{T}}(\varphi{}^{(b)}))^{H}
\end{align}
}where $\varphi{}^{(b)}$ is the sensing AoA/AoD, $\alpha^{(b)}$
is the reflection coefficient of the sensing channel gain, and $\gamma=\mathbb{E}\{\alpha^{(b)}\alpha^{(b)*}\}$.
The standard deviation of $\alpha^{(b)}$ is $\sqrt{\beta_{\mathrm{RCS}}(\lambda^{(b)})^{2}/(4\pi)^{3}(d^{(b)})^{4})}$,
where $d^{(b)}$ is the distance between the BS and sensing target,
and $\beta_{\mathrm{RCS}}$ is the \ac{RCS} \cite{dong2022sensing}.
Since the proposed optimization algorithm is applicable for any \ac{RCS},
we take $\beta_{\mathrm{RCS}}=1$ for ease of exposition. To enhance
numerical stability, we normalize the communication channel and the
target response matrices by the noise standard deviation, i.e., $\mathbf{\bar{H}}_{k}^{(b)}\leftarrow\mathbf{H}_{k}^{(b)}/\sigma^{(b)}$
and $\mathbf{\bar{G}}^{(b)}\leftarrow\mathbf{G}^{(b)}/\sigma^{(b)}$,
which implies $\mathbf{\bar{R}}^{(b)}\leftarrow\mathbf{R}^{(b)}/(\sigma^{(b)})^{2}$.

For radar signal processing, both $\widetilde{\mathbf{W}}^{(b)}$
and $\mathbf{\widetilde{S}}^{(b)}$ are known to the receiver. Thus,
the \ac{MI} between $\mathbf{y}_{S}^{(b)}$ and $\mathbf{g}^{(b)}$
can be calculated as \cite{li2024framework}{\small
\begin{gather}
\mathcal{I}(\mathbf{y}_{S}^{(b)};\mathbf{g}^{(b)}|\widetilde{\mathbf{W}}^{(b)},\mathbf{\widetilde{S}}^{(b)})\nonumber \\
=\mathcal{H}(\mathbf{y}_{S}^{(b)}|\widetilde{\mathbf{W}}^{(b)},\mathbf{\widetilde{S}}^{(b)})-\mathcal{H}(\mathbf{y}_{S}^{(b)};\mathbf{g}^{(b)}|\widetilde{\mathbf{W}}^{(b)},\mathbf{\widetilde{S}}^{(b)})\nonumber \\
=\ln\det(L\widetilde{\mathbf{W}}^{(b)}\mathbf{\bar{R}}^{(b)}(\widetilde{\mathbf{W}}^{(b)})^{H}+\mathbf{I}_{N_{\mathrm{tot}}N_{r}}).\label{eq:MI_sens}
\end{gather}
}From an information-theoretic point of view, we adopt the \ac{SR}
as the performance metric of sensing\footnote{\textcolor{black}{For the reason of consistency, the SR is chosen
as the performance metric for sensing since we use the CR as the metric
for communication. Alternatively, the MI can be used as the performance
metric for both communication and sensing, as was done in \cite{bazzi2025mutual}.}}, which is defined as the sensing MI per unit time \cite{ouyang2023mimo}.
For target sensing at band $b$, the SR is given by{\small
\begin{align}
R_{S}^{(b)} & =\tfrac{1}{L}B^{(b)}\mathcal{I}(\mathbf{y}_{S}^{(b)};\mathbf{g}^{(b)}|\widetilde{\mathbf{W}}^{(b)},\mathbf{\widetilde{S}}^{(b)})\nonumber \\
 & =\tfrac{1}{L}B^{(b)}\ln\det(L\widetilde{\mathbf{W}}^{(b)}\mathbf{\bar{R}}^{(b)}(\widetilde{\mathbf{W}}^{(b)})^{H}+\mathbf{I}_{N_{\mathrm{tot}}N_{r}}).\label{eq:SensRate}
\end{align}

}Note that both the \ac{CR} in (\ref{eq:Rck}) and the \ac{SR}
in (\ref{eq:SensRate}) are expressed in nats/s for mathematical convenience.\vspace{-5mm}

\subsection{Problem Formulation}

In this paper, we are interested in designing a cooperative ISAC system
that maximizes the \ac{SR} constrained by the total transmit power
budget and the per-user \ac{CR} requirements, leading to the following
optimization problem:{\small 
\begin{subequations}
\label{eg:Orig_prob}
\begin{align}
\text{\ensuremath{\underset{\{\mathbf{W}_{k}^{(b)}\}}{\maximize}}	} & \ \sum\nolimits_{b\in\mathcal{B}}R_{S}^{(b)}\label{eq:Obj_fun}\\
\st & \ \sum\nolimits_{b\in\mathcal{B}}\sum\nolimits_{k=1}^{K}\tr(\mathbf{W}_{k}^{(b)}(\mathbf{W}_{k}^{(b)})^{H})\le P_{\text{max}},\label{eq:Pow_budget}\\
 & \ \sum\nolimits_{b\in\mathcal{B}}R_{C,k}^{(b)}\ge r_{\min},\forall k\label{eq:Rc_min}
\end{align}
\end{subequations}
 }where (\ref{eq:Pow_budget}) ensures the total transmit power does
not exceed the maximum allowable power, and (\ref{eq:Rc_min}) guarantees
each user achieves at least the minimum required \ac{CR} $r_{\min}$.
Since we do not consider complex signal combining of the echo signals
across different frequency bands, the sum \ac{SR} serves as an appropriate
objective function in (\ref{eg:Orig_prob}).\vspace{-5mm}

\section{Proposed Solution}

Since solving for $\{\mathbf{W}_{k}^{(b)}\}$ directly is challenging,
we resort to the semi-definite rank relaxation method by introducing
the covariance matrices $\mathbf{Q}_{k}^{(b)}=\mathbf{W}_{k}^{(b)}(\mathbf{W}_{k}^{(b)})^{H}$.
As a result, the \ac{SR} at band $b$ can be rewritten as{\small
\begin{align}
\!\!\!R_{S}^{(b)} & =\tfrac{1}{L}B^{(b)}\ln\det(L\widetilde{\mathbf{W}}^{(b)}\mathbf{\bar{R}}^{(b)}(\widetilde{\mathbf{W}}^{(b)})^{H}+\mathbf{I}_{N_{\mathrm{tot}}N_{r}})\nonumber \\
 & \!\!\!\!\!\!\!\!\!\!\!\!=\tfrac{1}{L}B^{(b)}\ln\det\bigl[L\bigl(\mathbf{I}_{N_{r}}\otimes\sum\nolimits_{k=1}^{K}\mathbf{Q}_{k}^{(b)}\bigr)\mathbf{\bar{R}}^{(b)}+\mathbf{I}_{N_{t}N_{r}}\bigr]
\end{align}
}where we used the identities $\ensuremath{\det}(\ensuremath{\mathbf{X}}\ensuremath{\mathbf{Y}}+\mathbf{I}_{m})=\ensuremath{\det}(\ensuremath{\mathbf{Y}}\ensuremath{\mathbf{X}}+\mathbf{I}_{n})$,
$(\ensuremath{\mathbf{X}}_{1}\ensuremath{\mathbf{Y}}_{1})\otimes(\ensuremath{\mathbf{X}}_{2}\ensuremath{\mathbf{Y}}_{2})=(\ensuremath{\mathbf{X}}_{1}\otimes\ensuremath{\mathbf{X}}_{2})(\ensuremath{\mathbf{Y}}_{1}\otimes\ensuremath{\mathbf{Y}}_{2})$
and (\ref{eq:Sapprox}). We remark that $R_{S}^{(b)}$ is jointly
concave with all $\mathbf{Q}_{k}^{(b)}$, which makes the objective
function in (\ref{eq:Obj_fun}) easier to handle. Similarly, the \ac{CR}
of user $k$ at band $b$ is given by{\small
\begin{multline}
\!\!\!R_{C,k}^{(b)}=B^{(b)}\bigl[\ln\det(\mathbf{B}_{k}^{(b)}+\mathbf{\bar{H}}_{k}^{(b)}\mathbf{Q}_{k}^{(b)}\mathbf{\bar{H}}_{k}^{(b)H})-\ln\det(\mathbf{B}_{k}^{(b)})\bigr]\label{eq:Rqck}
\end{multline}
}where $\mathbf{B}_{k}^{(b)}=\sum_{j=1,j\neq k}^{K}\mathbf{\bar{H}}_{k}^{(b)}\mathbf{Q}_{j}^{(b)}\mathbf{\bar{H}}_{k}^{(b)H}+\mathbf{I}_{N_{k}}$.
Therefore, the original optimization problem (\ref{eg:Orig_prob})
can be equivalently reformulated as
\begin{subequations}
\label{eg:Orig_prob:SDP:relax}{\small
\begin{align}
\text{\ensuremath{\underset{\mathbf{Q}_{k}^{(b)}\in\mathcal{Q}}{\maximize}}	} & \ f(\{\mathbf{Q}_{k}^{(b)}\})\triangleq\sum\nolimits_{b\in\mathcal{B}}R_{S}^{(b)}\label{eq:Obj_fun-1}\\
\st & \ \sum\nolimits_{b\in\mathcal{B}}R_{C,k}^{(b)}\ge r_{\min},\forall k\label{eq:Rc_min-1}\\
 & \ \mathrm{rank}(\mathbf{Q}_{k}^{(b)})=N_{k}\label{eq:Rank_cons}
\end{align}
}
\end{subequations}
where $\mathcal{Q}\triangleq\{\mathbf{Q}_{k}^{(b)}\,|\,\sum\nolimits_{b\in\mathcal{B}}\sum\nolimits_{k=1}^{K}\tr(\mathbf{Q}_{k}^{(b)})\le P_{\text{max}}\}$
and (\ref{eq:Rank_cons}) ensures that $\mathbf{W}_{k}^{(b)}$ can
be exactly computed from $\mathbf{Q}_{k}^{(b)}$.

It is straightforward to see that the primary challenges in solving
(\ref{eg:Orig_prob:SDP:relax}) are the non-convex constraints (\ref{eq:Rc_min-1})
and (\ref{eq:Rank_cons}). To overcome these, we drop the rank constraint
(\ref{eq:Rank_cons}), and then apply the IA method to solve the resulting
problem \cite{marks1978general}. To this end, a concave lower bound
for $R_{C,k}^{(b)}$ is required. We note that $R_{C,k}^{(b)}$ is
in fact the difference of two concave functions. Thus, a lower bound
can be easily obtained by linearizing the term $\ln\det(\mathbf{B}_{k}^{(b)})$.
Let $\mathbf{Q}_{k}^{(b),n}$ and $\mathbf{B}_{k}^{(b),n}$ be the
values of $\mathbf{Q}_{k}^{(b)}$ and $\mathbf{B}_{k}^{(b)}$ in the
$n$-th iteration, respectively. Then the following inequality holds:{\small
\begin{multline}
\ln\det(\mathbf{B}_{k}^{(b)})\le\ln\det(\mathbf{B}_{k}^{(b),n})+\\
\sum\nolimits_{i=1,i\neq k}^{K}\tr\bigl(\mathbf{H}_{k}^{(b)H}(\mathbf{B}_{k}^{(b),n})^{-1}\mathbf{H}_{k}^{(b)}(\mathbf{Q}_{i}^{(b)}-\mathbf{Q}_{i}^{(b),n})\bigr)\label{eq:approx_term}
\end{multline}
}which is due to the concavity of the $\ln\det()$ function.

Substituting (\ref{eq:approx_term}) into (\ref{eq:Rqck}), a lower
bound on the \ac{CR} is given by{\small
\begin{gather}
R_{C,k}^{(b)}\ge\bar{R}_{C,k}^{(b)}=B^{(b)}\Bigl[\ln\det(\mathbf{B}_{k}^{(b)}+\mathbf{\bar{H}}_{k}^{(b)}\mathbf{Q}_{k}^{(b)}\mathbf{\bar{H}}_{k}^{(b)H})-\ln\det(\mathbf{B}_{k}^{(b),n})\nonumber \\
-\sum_{i=1,i\neq k}^{K}\tr\bigl(\mathbf{\bar{H}}_{k}^{(b)H}(\mathbf{B}_{k}^{(b),n})^{-1}\mathbf{\bar{H}}_{k}^{(b)}(\mathbf{Q}_{i}^{(b)}-\mathbf{Q}_{i}^{(b),n})\bigr)\Bigr].\label{eq:Rq_LB}
\end{gather}
}which is jointly concave with respect to all $\mathbf{Q}_{k}^{(b)}$.
Utilizing (\ref{eq:Rq_LB}), $\mathbf{Q}_{i}^{(b),n+1}$ is obtained
as the solution to the following convex problem:
\begin{subequations}
\label{eg:Orig_subproblem}{\small
\begin{align}
\text{\ensuremath{\underset{\mathbf{Q}_{k}^{(b)}\in\mathcal{Q}}{\maximize}}	} & \quad f(\{\mathbf{Q}_{k}\})\label{eq:Obj_fun-1-2}\\
\st & \quad\sum\nolimits_{b\in\mathcal{B}}\bar{R}_{C,k}^{(b)}\ge r_{\min},\forall k,\label{eq:Rc_min_approx}
\end{align}
}
\end{subequations}

\paragraph*{Initialization}

The proposed method requires a feasible point to start, which is not
trivial to obtain due to the nonconvexity of $R_{C,k}^{(b)}$. To
overcome this issue, we consider the following regularized problem
of (\ref{eg:Orig_subproblem}):
\begin{subequations}
\label{eg:subproblem:regularized}{\small
\begin{align}
\text{\ensuremath{\underset{\mathbf{Q}_{k}^{(b)}\in\mathcal{Q},s_{k}\geq0}{\maximize}}	} & \quad f(\{\mathbf{Q}_{k}\})-\rho\sum\nolimits_{k=1}^{K}s_{k}\label{eq:obj_regularized}\\
\st & \quad\sum\nolimits_{b\in\mathcal{B}}\bar{R}_{C,k}^{(b)}+s_{k}\ge r_{\min},\forall k\label{eq:Rc_min-1-1-1}
\end{align}
}
\end{subequations}
where $s_{k}\geq0$ ($k=1,2,\dots,K$) are slack variables, and $\rho>0$
is the regularization parameter. It is straightforward to see that
(\ref{eg:subproblem:regularized}) is always feasible for any initial
choice of $\mathbf{Q}_{i}^{(b),0}$. Due to the regularization term,
the slack variables are gradually forced to zero as the iterations
progress. Once this happens, the obtained $\mathbf{Q}_{k}^{(b)}$
is feasible to (\ref{eg:Orig_prob:SDP:relax}), and thus, can be used
to initialize the proposed iterative optimization method. \algref{proposed_IA}
outlines the overall proposed method.\SetAlFnt{\small} 
\begin{algorithm}[t]
\caption{Optimization of the covariance matrices.\label{alg:proposed_IA}}

\SetAlgoNoLine
\DontPrintSemicolon
\LinesNumbered 

\KwIn{$\mathbf{Q}_{k}^{(b),0}$ (randomly generated), $\rho\ge0$,
$m\leftarrow0$}

\Repeat(\tcp*[f]{find a feasible point}){$s_{k}<\epsilon\:(\forall k)$}{

\hspace*{-10pt}Solve (\ref{eg:subproblem:regularized}) to obtain
$\mathbf{Q}_{k}^{(b),m+1}$ and increase $m$: $m\leftarrow m+1$\;

}

Reset $\mathbf{Q}_{k}^{(b),0}\leftarrow\mathbf{Q}_{k}^{(b),m+1}$
and set $n\leftarrow0$ \label{alg:feasible:achieved}

\Repeat{convergence}{

\hspace*{-10pt}Solve (\ref{eg:Orig_subproblem}) to obtain $\mathbf{Q}_{k}^{(b),n+1}$
and increase $n$: $n\leftarrow n+1$ \;

}

\KwOut{ $\mathbf{Q}_{k}^{(b),\ast}=\mathbf{Q}_{k}^{(b),n+1},k=1,2,\ldots,K$
}
\end{algorithm}

\begin{rem}
After the optimization is complete, the appropriate precoding matrices
$\mathbf{W}_{k}^{(b)}$ can be obtained from the covariance matrices
$\mathbf{Q}_{k}^{(b),\ast}$. If $\text{rank}(\mathbf{Q}_{k}^{(b),\ast})=N_{k}$,
then $\mathbf{W}_{k}^{(b)}$ can be obtained exactly from $\mathbf{Q}_{k}^{(b),\ast}$
by \ac{EVD}. If not, then randomization methods introduced in \cite{sidiropoulos2006transmit}
are applied to find $\mathbf{W}_{k}^{(b)}$. However, in our extensive
numerical experiments, the condition $\text{rank}(\mathbf{Q}_{k}^{(b),\ast})=N_{k}$
is always met, and thus, randomization is not necessary. This strongly
suggests the potential existence of an analytical proof for this empirical
observation. Pursuing such a proof is beyond the scope of this work,
and we leave this interesting open problem for future research. For
the high-frequency bands, $\mathbf{W}_{BB,k}^{(b)}$ and $\mathbf{W}_{RF,k}^{(b)}$
are obtained from $\mathbf{W}_{k}^{(b)}$ by applying \ac{OMP} \cite[Algorithm 1]{el2014spatially}.
\end{rem}

\paragraph*{\textcolor{black}{Complexity Analysis}}

\textcolor{black}{To analyze the computational complexity for solving
(\ref{eg:subproblem:regularized}), we have to transform it into a
more standard form, which includes positive semidefinite (PSD) and
exponential cones. To get a crude estimate of the complexity, we neglect
the exponential cones, since the computational complexity is mostly
dominated by the PSD cones and can be approximated by $\mathcal{O}(B^{3.5}N_{r}^{6.5}N_{t}^{6.5})$
\cite[Section 6.6.3]{ben2001lectures}.}

\paragraph*{Convergence Analysis}

The convergence analysis of the proposed method closely follows the
arguments presented in \cite{marks1978general}. It is easy to see
that the inequality in (\ref{eq:Rq_LB}) is tight when $\mathbf{Q}_{i}^{(b)}=\mathbf{Q}_{i}^{(b),n}$,
$i=1,2,\ldots,K$, which implies that $\{\mathbf{Q}_{i}^{(b),n}\}$
is feasible to (\ref{eg:Orig_subproblem}). Since $\{\mathbf{Q}_{i}^{(b),n+1}\}$
is the global solution to (\ref{eg:Orig_subproblem}), it follows
that $f(\{\mathbf{Q}_{k}^{n+1}\})\geq f(\{\mathbf{Q}_{k}^{n}\})$,
i.e., the objective sequence $f(\{\mathbf{Q}_{k}^{n}\})$ is nondecreasing.
Assuming that the feasible set of (\ref{eg:Orig_prob:SDP:relax})
is nonempty, then the sequence $f(\{\mathbf{Q}_{k}^{n}\})$ is bounded
above, and is thus convergent since the feasible set is compact.

We remark that the objective function in (\ref{eq:obj_regularized})
strikes a balance between maximizing the \ac{SR} and ensuring the
feasibility of (\ref{eg:Orig_subproblem}). Thus, once a feasible
starting point is achieved (cf. Line \ref{alg:feasible:achieved}
of \algref{proposed_IA}), it is also a good starting point. This,
in turn, accelerates the convergence of the main iterative procedure,
enabling the algorithm to terminate within only a few iterations,
as shall be demonstrated in the next section.

\section{Simulation Results}

In this section, we evaluate the \ac{SR} of the proposed algorithm
for a cooperative ISAC system through Monte Carlo simulations and
compare it with benchmark schemes. In the first benchmark scheme,
which is denoted as \emph{Eq-Pow-Split}, the total power budget is
uniformly distributed among all BSs, i.e., the power budget of each
BS is $P_{\text{max}}/B$. In the other three benchmark schemes, which
are denoted as \emph{BS1 Only}, \emph{BS2 Only} and \emph{BS3 Only},
only the BS1, BS2 and BS3 are active, respectively.

In the simulation setup, we consider three frequency bands $f^{(1)}=6\,\text{GHz}$
($\lambda^{(1)}=5\,\text{cm}$), $f^{(2)}=26\,\text{GHz}$ ($\lambda^{(2)}\approx1.15\,\text{cm}$),
$f^{(3)}=26.5\,\text{GHz}$ ($\lambda^{(3)}\approx1.13\,\text{cm}$)
with corresponding bandwidths $B^{(1)}=1\,\text{MHz}$ and $B^{(2)}=B^{(3)}=4\,\text{MHz}$.
That is, BS1 operates in a low-frequency band, while BS2 and BS3 operate
in a high-frequency bands. The number of propagation paths is $P^{(1)}=8$
and $P^{(2)}=P^{(3)}=4$, while the minimum communication rate requirement
is $r_{\min}=100\,\mathrm{kbits/s}$. Other parameters are set as
$N_{t}=8$, $N_{r}=2$, $N_{k}=2$, $P_{\max}=0.1\thinspace\mathrm{W}$,
$L=30$, $\rho=1$, $\epsilon=10^{-5}$. The noise power is calculated
according to $(\sigma^{(b)})^{2}=k_{B}TB^{(b)}F$, where $k_{B}=1.381\times10^{-23}$
is the Boltzmann constant, $T=290\,\text{K}$ is the standard temperature,
and $F=7.94$ (i.e., $9\,\text{dB}$) is the noise figure of the receiver.
The midpoints of the transmit and the receive \acp{ULA} for BS1,
BS2, and BS3 are at $(0,5\,\mathrm{m},0)$, $(0,5\,\mathrm{m},300\,\mathrm{m})$,
and $(210\,\text{\ensuremath{\mathrm{m}}},5\,\mathrm{m},-80\,\text{\ensuremath{\mathrm{m}}})$,
respectively, with \acp{ULA} at BS1 and BS2 oriented along the $z$-axis,
and at BS3 along the $x$-axis. The midpoint of the $k$-th user’s
ULA is $(25\,\mathrm{m},1.5\,\mathrm{m},z_{k})$, where $z_{k}$ is
uniformly distributed in $[25\,\mathrm{m},275\,\mathrm{m}]$, while
the point target is located at $(-25\,\mathrm{m},1\,\mathrm{m},z_{t})$,
where $z_{t}$ follows the same uniform distribution as $z_{k}$.
The inter-antenna spacing for each BS is half of the wavelength of
its operating frequency, while for users, this spacing is $\lambda^{(1)}/2$
to avoid antenna coupling. The AoA ($\theta_{p,k}^{(b)}$) and AoD
($\phi_{p,k}^{(b)}$) are uniformly sampled from $[-\pi/2,\pi/2]$.
The CVX tool with MOSEK as the internal software package is used to
implement \algref{proposed_IA}. The initial values of the covariance
matrices are randomly generated while satisfying the total power budget.
\algref{proposed_IA} terminates once the relative change of the SR
is less than $10^{-3}$. Moreover, the SR is measured in bits per
second (bits/s) and averaged over 500 independent channel realizations.
\begin{figure}[t]
\centering{}\subfloat[\textcolor{black}{For two BSs.}]{\centering{}\includegraphics[width=4.25cm]{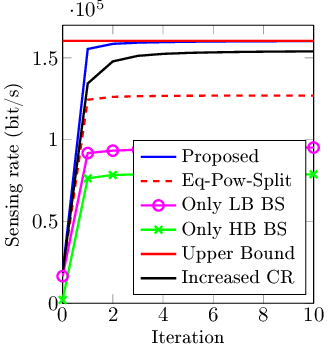}}\subfloat[For three BSs.\label{fig:3BSs.}]{\centering{}\includegraphics[width=4.25cm]{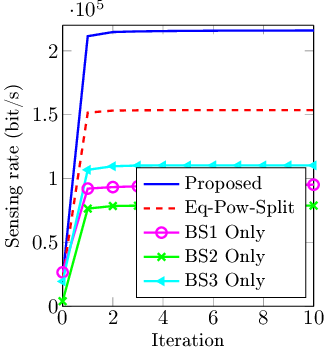}}\caption{Convergence of the proposed algorithm and of the benchmark schemes
for different numbers of BSs.\label{fig:Convergence}}
\end{figure}

In Fig. \ref{fig:Convergence}, we present the convergence behavior
of the proposed algorithm and of the benchmark schemes for a cooperative
ISAC system with two and three BSs. In general, \emph{all schemes
require only a few iterations to fully converge after a feasible point
is found}. Due to its ability to optimally distribute power among
the BSs, the proposed algorithm achieves the largest \ac{SR}, outperforming
the \emph{Eq-Pow-Split} scheme by approximately 25\,\% and 40\,\%
in cooperative ISAC systems with two and three BSs, respectively.
\textcolor{black}{In addition, the proposed scheme achieves the same
SR as the }\textcolor{black}{\emph{Upper Bound}}\textcolor{black}{{}
benchmark scheme, which excludes the CR constraint and therefore provides
an upper bound on the achievable SR. The SR of }\textcolor{black}{\emph{Upper
Bound}}\textcolor{black}{{} can be obtained by using the water-filling
algorithm. Furthermore, substantially increasing the minimum CR, $r_{\min}$,
to 24\,Mbit/s (i.e., the }\textcolor{black}{\emph{Increased CR}}\textcolor{black}{{}
benchmark scheme) causes a negligible reduction in the SR. The SRs
of }\textcolor{black}{\emph{Upper Bound}}\textcolor{black}{{} and }\textcolor{black}{\emph{Increased
CR}}\textcolor{black}{{} in a system with 3 BSs follow the same trend
as in a system with 2 BSs, and thus, are omitted from Fig. 1(b) for
the purpose of clarity. These results confirm the effectiveness of
the proposed algorithm}\textcolor{black}{\emph{.}} Regarding benchmark
schemes with only one active BS, \emph{BS3 Only} achieves the highest
individual \ac{SR}, although it operates at a higher frequency than
the other BSs. This can be attributed to the fact that the BS3 is
in most cases closer to the sensing target, particularly when the
target is equidistant from the first two BSs. These results indicate
that, besides the operating frequency, the distance from the sensing
target is another key factor that determines the \ac{SR} in ISAC
cooperative systems. In general, we can observe that the proposed
algorithm for cooperative ISAC systems generally offers a substantially
larger \ac{SR} compared to benchmark schemes with only one active
BS. Moreover, in our simulations, $\text{rank}(\mathbf{Q}_{k}^{(b),\ast})$
is always equal to $N_{k}$ for $k=1,2,\ldots,K$, which implies that
dropping the rank constraint does not affect the optimality of (\ref{eg:Orig_prob:SDP:relax}).

\begin{figure}[t]
\centering{}\includegraphics[width=8.85cm]{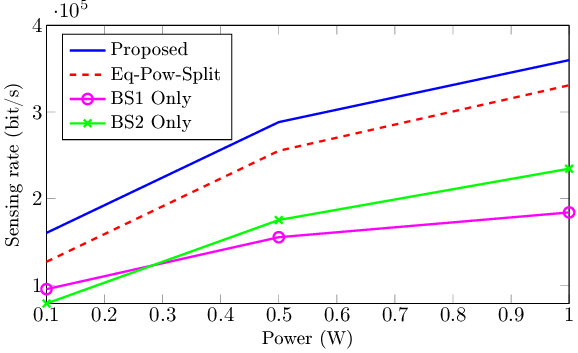}\caption{\textcolor{black}{Variations of the \ac{SR} with the transmit power.
\label{fig:SR-vs-P}}}
\end{figure}

The variation of the SR with the transmit power for the case of two
BSs is shown in Fig. \ref{fig:SR-vs-P}. Similar results were obtained
for the case of three BSs. As expected, the \acp{SR} of all schemes
increase logarithmically with the transmit power. The advantage of
using the cooperative ISAC capabilities of both BSs, rather than relying
on a single BS, becomes more apparent with higher transmit power.
The proposed scheme consistently achieves a larger \ac{SR} than \emph{Eq-Pow-Split},
but the performance gap remains approximately constant across different
power levels. This suggests that the proposed scheme is particularly
beneficial in the low to medium transmit power range, while \emph{Eq-Pow-Split
}serves as an effective sub-optimal solution in the high transmit
power regime. Regarding the benchmark schemes with only one active
BS, \emph{BS1 Only} achieves a larger \ac{SR} in the lower transmit
power region, while the opposite is true in the higher transmit power
region. This is due to the fact that at higher transmit power levels,
the higher propagation losses associated with higher-frequency bands
(e.g., BS2) can be compensated by increased channel directivity, which
enhances sensing performance.\vspace{-3mm}

\section{Conclusion}

In this paper, we have studied the sum \ac{SR} maximization in a
cooperative ISAC system with linear precoding, where each BS operates
in a different frequency band. With this aim, we developed an algorithm
based on the semi-definite rank relaxation method by introducing the
covariance matrices of the appropriate precoding matrices as optimization
variables and utilized the IA method to deal with the nonconvexity
of the resulting problem. Simulation results show that the proposed
algorithm increases the SR by approximately 25\,\% and 40\,\% compared
to the case of equal power distribution in a cooperative ISAC system
with two and three BSs, respectively. Additionally, the algorithm
converges in a couple of iterations. Lastly, we conclude that the
proposed method is most effective in the low transmit power regime,
where power-efficient cooperative sensing and communication are essential\textcolor{black}{.
Future work includes the evaluation of the proposed method under more
advanced target models, including the incorporation of a clutter model.}%
\begin{comment}
\begin{itemize}
\item You need to use the journal abbreviations. Check the file IEEEabrv
and update the .bib file accordingly. For example: journal=\{IEEE
Transactions on Vehicular Technology\} -> journal=IEEE\_J\_VT
\item For well-known conferences, use abbreviations as well to save space.
2024 IEEE Wireless Communications and Networking Conference (WCNC)
-> IEEE WCNC
\item Abbreviations in the papers title should be written in capital: ``Two-timescale
design for ap mode selection of cooperative isac networks'' -> Two-timescale
design for AP mode selection of cooperative ISAC networks.
\item To save some space, you can for Bibtex to write first author et al.
if there are more than 2 authors.
\end{itemize}
\end{comment}
\vspace{-3mm}\bibliographystyle{IEEEtran}
\bibliography{IEEEabrv,Cooperative_ISAC_final}

\end{document}